\documentclass[lettersize,journal]{IEEEtran}
\usepackage{amsmath,amsfonts}
\usepackage{algorithmic}
\usepackage{algorithm}
\usepackage{array}
\usepackage[caption=false,font=normalsize,labelfont=sf,textfont=sf]{subfig}
\usepackage{textcomp}
\usepackage{stfloats}
\usepackage{url}
\usepackage{verbatim}
\usepackage{graphicx}
\usepackage{cite}
\usepackage[separate-uncertainty=true,multi-part-units=single]{siunitx}
\usepackage{adjustbox} 
\usepackage{mathrsfs}
\usepackage{multirow}
\usepackage{booktabs}
\usepackage{mhchem}
\newcommand{\ie}{\textit{i}.\textit{e}., }
\newcommand{\etal}{\textit{et al}.}
\begin{document}

\title{Spectral Thermal Spreading Resistance of Wide Bandgap Semiconductors in Ballistic-Diffusive Regime}

\author{Yang Shen, Yu-Chao Hua, Han-Ling Li, S.L. Sobolev, Bing-Yang Cao
	\thanks{This work was supported by the National Natural Science Foundation of China (Nos. 51825601,U20A20301). The reported study was also funded by RFBR and NSFC (Nos. 20-58-53017, 52011530030). (Corresponding author: Bing-Yang Cao.)}
	\thanks{Yang Shen, Han-Ling Li, Bing-Yang Cao are with the Key Laboratory of Thermal Science and Power Engineering of Education of Ministry, Department of Engineering Mechanics,
	Tsinghua University, Beijing 100084, China (e-mail: sy980829@163.com; lihanling1994@163.com; caoby@tsinghua.edu.cn).}
	\thanks{Yu-Chao Hua is with the LTEN laboratory, Polytech Nantes, University of Nantes, UMR6607, F-44000, Nantes, France (e-mail: huayuchao19@163.com).}
	\thanks{S.L. Sobolev is with 
the Institute of Problems of Chemical Physics, Academy of Sciences of Russia, Chernogolovka, Moscow Region, 142432, Russia and Samara State Technical University, ul. Molodogvardeiskaya 244, Samara, 443100, Russia (e-mail: sobolev@icp.ac.ru).}}



\maketitle

\begin{abstract}
	To develop efficient thermal management strategies for wide bandgap (WBG) semiconductor devices, it is essential to have a clear understanding of the heat transport process within the device and accurately predict the junction temperature. In this paper, we used the phonon Monte Carlo (MC) method with the phonon dispersion of various typical WBG semiconductors, including GaN, SiC, AlN, and \ce{\beta-Ga_2O_3}, to investigate the thermal spreading resistance in a ballistic-diffusive regime. It was found that when compared with Fourier's law-based predictions, the increase in the thermal resistance caused by ballistic effects was strongly related to different phonon dispersions. Based on the model deduced under the gray-medium approximation and the results of dispersion MC, we obtained a thermal resistance model that can well address the issues of thermal spreading and ballistic effects, and the influences of phonon dispersion. The model can be easily coupled with FEM based thermal analysis and applied to different materials. This paper can provide a clearer understanding of the influences of phonon dispersion on the thermal transport process, and it can be useful for the prediction of junction temperatures and the development of thermal management strategies for WBG semiconductor devices.
\end{abstract}


\section{Introduction}

Wide bandgap (WBG) semiconductor devices such as GaN high electron mobility transistors (HEMTs) \cite{mishra2002algan,zeng2018comprehensive}, SiC metal-semiconductor field-effect transistors (MESFETs) \cite{sriram2009high,zhu2019improved} and recently burgeoning AlN and \ce{\beta-Ga_2O_3} based ultra-WBG (UWBG) devices \cite{higashiwaki2012gallium,wong2015field,wang2021progresses} are extremely attractive for high-power and high-frequency electronic applications, due to the high breakdown voltage led by the wide bandgap. However, owing to their superhigh power density, they usually hold very high junction temperatures and the significant overheating within the devices largely restricts their actual performance \cite{wu2007transient, cheng2019significantly, liu2020machine} and shortens the device lifetime \cite{waltereit2012influence}. Therefore, it is important to have a clear understanding of the heat transport process within the devices, to accurately predict the junction temperature and thereby develop effective thermal management strategies \cite{bagnall2013device,tang2020thermal,tang2021phonon}.

In general, there are two prominent features of the heat transport process within WBG semiconductor devices. First, the heat generation is extremely localized at the top of the channel layer \cite{chatterjee2019device, chatterjee2020nanoscale}. The heat source region is extremely small \cite{sarua2007thermal} compared with the channel layer length and width. When heat spreads from a small heat source region to a much larger region, there is a significant thermal spreading resistance, which is one of the main sources of the total thermal resistance \cite{razavi2016review} and finally results in a large near-junction temperature spike \cite{muzychka2013thermal}. Second, the basic structure of the device is made up of multilayer micro/nano films, and the thickness of the channel layer is usually less than \SI{3}{\um} \cite{chatterjee2020nanoscale, garven2009simulation, orouji2011novel}. This characteristic size is comparable with the mean free paths (MFPs) of phonons, which are the main heat carriers in semiconductors \cite{chen2021non}. The transport of phonons in the device is significantly suppressed by strong phonon-boundary scatterings, leading to an additional ballistic thermal resistance \cite{hua2016ballistic} and further increasing the juncture temperature \cite{hua2019thermal}. In this case, Fourier's law becomes inapplicable and ballistic-diffusive heat conduction emerges\cite{bao2018review}. Therefore, to accurately predict the junction temperature, it is important to provide a clear understanding of the thermal spreading process and the influences of ballistic effects.

Although great efforts have been devoted to multiscale simulation techniques to predict the temperature profile of devices \cite{sadi2010monte, hao2017hybrid, hao2018hybrid}, little work has analyzed the thermal spreading process and the influences of ballistic effects in detail \cite{hua2019thermal}. In addition, these simulations often take a long time to converge, and when considering phonon dispersion, they can be much more time-consuming \cite{li2020ballistic}. The high computational demands partly restrict the applications of these simulation techniques in the practical 3D thermal analysis and optimizations of real devices, where the finite-element method (FEM) still plays a major role \cite{wang2013simulation, zhang2019enhancement,yuan2020modeling}. Therefore, it is still urgent and significant to propose new physical insights into the thermal transport process within the device and thereby develop computationally efficient models that can take multiple factors into account and can give a fast estimation of the junction temperature. Additionally, the model should be easily coupled with the widely used FEM-based design process. 

Based on Fourier's heat conduction law, analytical thermal spreading resistance models have been extensively studied \cite{razavi2016review,muzychka2013thermal,bagnall2014analytical}. Recently, Hua \etal  \cite{hua2019thermal} developed a semiempirical thermal resistance model that could consider the influences of both thermal spreading and ballistic effects under the gray-medium approximation. The results indicated that ballistic effects could significantly alter the temperature distributions within channel layers and can greatly increase the thermal resistance. However, this work did not scrutinize the influences of phonon dispersion. Since thermal transport in semiconductors is indeed the contribution of all phonon modes, different phonon dispersion relations will significantly affect the juncture temperature \cite{li2020ballistic}. For typical WBG semiconductors in particular, Freedman \etal \cite{freedman2013universal} found that phonons with MFPs greater than \SI{1000(230)}{\nm}, \SI{2500(800)}{\nm}, and \SI{4200(850)}{\nm} contribute 50\% of the bulk thermal conductivity of GaN, AlN and 4H-SiC near room temperature, respectively. In consideration of the characterized size of the channel layer, the phonon transport in the device should undergo strong size effects caused by the wide distributions of phonon MFPs, and it is necessary to reexamine the applicability of the semiempirical gray model and analyze the impacts of phonon dispersion in more detail.

In this work, we used phonon MC methods with phonon dispersion of various typical WBG semiconductors to study the thermal spreading resistance in ballistic-diffusive regimes. The considered materials include GaN, SiC, AlN and \ce{\beta-Ga_2O_3}. It was found that the gray model was insufficient to reflect the strong ballistic effects caused by the wide phonon MFP distributions. Based on the analysis of the deviations, material-independent correction factors were introduced to the gray model to better reflect the influences of phonon dispersion on the thermal transport process. The improved model can well address the issues of thermal spreading, ballistic effects, and the influences of phonon dispersion, and it is easy to couple with FEM-based thermal analysis. Our work could be helpful for the junction temperature predictions and the development of thermal management strategies for WBG semiconductors. 

\section{Problem Formulation and Methodologies}

\subsection{Problem Statement}

\begin{figure}[htbp]
	\centering
	\includegraphics[width=0.95\linewidth]{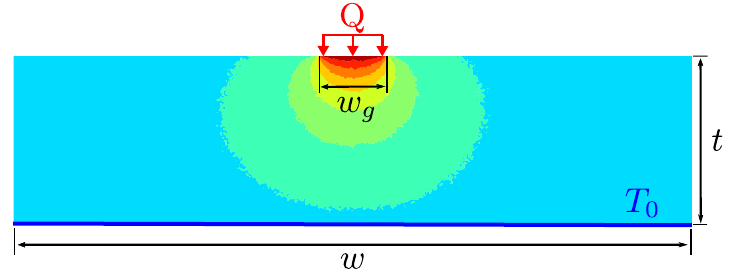}
	\caption{Schematic for the basic system for thermal spreading resistance analysis.}
	\label{fig:schematic}
\end{figure}

Fig. \ref{fig:schematic} shows the basic system for thermal spreading resistance analysis, which is a representative 2D simplified model for the channel layer. The geometry of the system can be represented by three parameters: the channel thickness $t$, the width of the structure $w$, and the width of the heating area $w_g$. At the top surface of the structure, uniform heat flux is specified over the region of the heat source, and the total heating power is denoted by $Q$. The remainder of the heat source surface is taken as adiabatic. The lateral boundaries are set as periodic, and the bottom boundary serves as an isothermal heat sink.
    
The total thermal resistance can be calculated by using the mean temperature over the heat source area ($\bar{T_s}$), heat sink temperature ($T_0$), and total heating power ($Q$) \cite{razavi2016review}, 
\begin{equation}
	R_t = \frac{\bar{T_s} - T_0}{Q}.
\end{equation}
For convenience in identifying the thermal resistance model \cite{hua2019thermal}, the dimensionaless total thermal resistance is defined as $R_t / R_{1d\_0}$. $R_{1d\_0}$ is the one-dimensional thermal resistance of purely diffusive heat conduction,
\begin{equation}
	R_{1d\_0} = t / (wk_0),
\end{equation}
where $k_0$ is the intrinsic thermal conductivity.

\subsection{Semiempirical Thermal Resistance Model}
\label{subsec:model}

For the system shown in Fig. \ref{fig:schematic}, the total thermal resistance was rearranged to \cite{hua2019thermal}
\begin{equation}
	\frac{R_t}{R_{1d\_0}} = \frac{R_F}{R_{1d\_0}}\cdot  \frac{R_{1d}}{R_{1d\_0}}\cdot  \left[ \frac{R_t}{R_{1d}}\left( \frac{R_F}{R_{1d\_0}} \right)^{-1} \right],
\end{equation}
where
\begin{align}
	  & \frac{R_F}{R_{1d\_0}} = 1 + (\frac{w}{w_g})^2(\frac{w}{t})\sum_{n=1}^{\infty} \frac{8 \sin ^{2}\left(\frac{w_g n\pi}{2w}\right) \cdot \cos^2(\frac{n\pi}{2})}{(n\pi)^3 \coth(\frac{tn\pi}{w} )}, \label{eq:model_fourier} \\
	  & \frac{R_{1d}}{R_{1d\_0}} = 1 + \frac{2}{3} Kn_t, \label{eq:model_cross}                                                                                                                                                   \\
	  & \frac{R_t}{R_{1d}}\left( \frac{R_F}{R_{1d\_0}} \right)^{-1} = r_w = 1 + A_w\left( w_g/w, w/t \right)Kn_w \label{eq:model_source}.                                                                                         
\end{align}

$R_F / R_{1d\_0}$ is the analytical expression of the dimensionless total thermal resistance derived based on Fourier's law \cite{muzychka2003thermal}, which is used to characterize the thermal spreading effect. The remaining two terms are used to reflect the ballistic thermal resistance. To characterize the strength of ballistic effects, two Knudsen numbers $Kn_t$ and $Kn_w$ were defined as
\begin{align}
	Kn_t & = l_0/t, \label{eq:knudsen_t}     \\
	Kn_w & = l_0 / w_g, \label{eq:knudsen_w} 
\end{align}
where $l_0$ is the intrinsic phonon MFP. $R_{1d} / R_{1d\_0}$ can be derived with the differential approximation and the temperature jump boundary conditions \cite{hua2017slip}, which represents the cross-plane ballistic thermal resistance. The last term denoted by $r_w$ is used to reflect the ballistic effect with $w_g$ comparable with MFP and have a similar form with the expression of $R_{1d} / R_{1d\_0}$. $A_w$ is a fitting parameter that is a function of geometric factors $w_g/w$ and $w/t$.  

The model was derived under the gray-medium approximation, \ie for single phonon mode. When considering phonon dispersion, the size effects on different phonon modes should be considered separately. In this case, the effective thermal conductivity can be evaluated by using mode-dependent modified MFPs,
\begin{equation}
	k_{eff} = \frac{1}{3} \sum_{j} \int_{0}^{\omega_{j}} \hbar \omega \frac{\partial f_{0}}{\partial T} \operatorname{DOS}_{j}(\omega) v_{g \omega j} l_{j, m} \mathrm{d} \omega .
	\label{eq:conductivity}
\end{equation}
Where
\begin{equation}
	l_{j, m} = \frac{l_{0, j}}{\left(1+A_{w} K n_{w_{-} \omega, j}\right)\left(1+\frac{2}{3} K n_{t_{-} \omega, j}\right)},
	\label{eq:depressed_mfp}
\end{equation}
in which $l_{0,j}$ is the intrinsic frequency-dependent MFP of the $j$ phonon branch, $K n_{w_{-} \omega, j} = l_{0,j} / w_g$ and $K n_{t_{-} \omega, j} = l_{0,j}/t$ are the frequency-dependent and phonon branch-dependent Knudsen numbers. With the effective thermal conductivity, the total thermal resistance can be calculated with Fourier's law-based model as equation \eqref{eq:model_fourier}.

\subsection{Phonon Monte Carlo Simulations}
 
In this paper, the phonon tracing MC technique was used to simulate ballistic-diffusive heat conduction \cite{tang2016phonon,hua2014phonon}. The basic simulation settings are the same as those in Ref. \cite{hua2019thermal}. The main difference is that when considering phonon dispersion, the emitted phonon bundles will be sampled from the phonon spectrum, and their properties will be redetermined after phonon-phonon scattering \cite{li2020ballistic}. In this work, the energy-based variance-reduced technique proposed by \cite{peraud2011efficient} was adopted to determine the probability of drawing a phonon bundle at a certain polarization and frequency. Details of dispersion MC can be found in Ref. \cite{li2020ballistic}. 

Four typical WBG or UWBG semiconductors were considered in this work, including GaN, AlN, SiC and \ce{\beta-Ga_2O_3}. An isotropic sine-shaped phonon dispersion was used for all materials \cite{chung2004role}, $\omega(k)=\omega_{\max } \sin \left(\pi k / 2 k_{m}\right)$, where $k_{m}=\left(6 \pi^{2} n\right)^{1 / 3}$ with $n$ as the volumetric density of primitive cells. Relaxation time is also needed by dispersion MC to determine the MFPs of different phonon modes. The essential phonon scattering mechanisms include impurity scattering ($I$) and Umklapp phonon-phonon scattering ($U$). The relaxation time can be expressed as $\tau_{I}^{-1} = A\omega^4$ and $\tau_{U}^{-1} = B\omega^2T\exp(-C/T)$ \cite{chen2005nanoscale}, where $A, B,$ and $C$ are the fitting constants. The total relaxation time can be calculated using Matthiessen's rule $\tau^{-1} = \tau_{I}^{-1} + \tau_{U}^{-1}.$

\begin{figure}[htbp]
	\centering
	\includegraphics[width=0.8\linewidth]{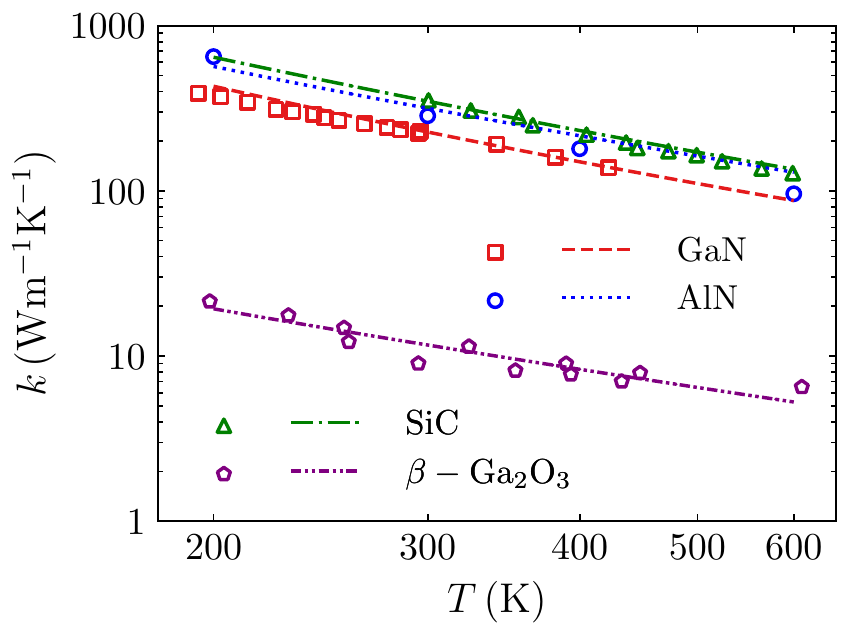}
	\caption{Thermal conductivity from the model calculations (line) , and from experiments (symbols) of GaN \cite{danilchenko2007upper}, AlN \cite{slack1987intrinsic}, 6H-SiC \cite{zheng2019thermal}, and \ce{\beta-Ga_2O_3} \cite{guo2015anisotropic, jiang2018three}.}
	\label{fig:conductivity}
\end{figure}

Parameters of different materials can be obtained by fitting the measured thermal conductivities. Table \ref{tab:dispersion_parameters} lists the parameters used in dispersion MC in this work. For GaN, AlN, and SiC, the fitted values in Ref. \cite{hao2017hybrid} were used. For \ce{\beta-Ga_2O_3}, the parameters were fitted using the thermal conductivities along the $[100]$ crystallographic direction  \cite{guo2015anisotropic,yan2018phonon,jiang2018three}. As shown in Fig. \ref{fig:conductivity}, the dispersion and fitted relaxation time model could well reflect the variance of the thermal conductivity with temperatures of different materials. 

\begin{table}[htbp]
	\centering
	\caption{Phonon dispersion and relaxation time parameters for different semiconductor materials.}
	\label{tab:dispersion_parameters}
	\begin{tabular}{@{}ccccc@{}}
		\toprule
		Parameter (Unit)                 & GaN   & AlN   & SiC   & $\ce{\beta-Ga_2O_3}$ \\ \midrule
		$k_0 \, (\SI{1e9}{m^{-1}})$      & 10.94 & 11.19 & 8.94  & 6.74                 \\
		$\omega_m \, (\SI{1e13}{rad/s})$ & 3.50  & 5.18  & 7.12  & 1.6                  \\
		$a_D\, (\si{\angstrom})$         & 2.87  & 2.81  & 3.51  & 4.66                 \\
		$A\, (\SI{1e-45}{s^3})$          & 5.26  & 10.5  & 1.00  & 1.38E-6              \\
		$B\, (\SI{1e-19}{s/K})$          & 1.10  & 0.728 & 0.596 & 9.31                 \\
		$C\, (\si{K})$                   & 200   & 287.5 & 235.0 & 62.6                 \\ \bottomrule
	\end{tabular}
\end{table}

\section{Results and Discussion}

\subsection{Total Thermal Resistance}

\begin{figure}[htbp]
	\centering
	\includegraphics[width=0.8\linewidth]{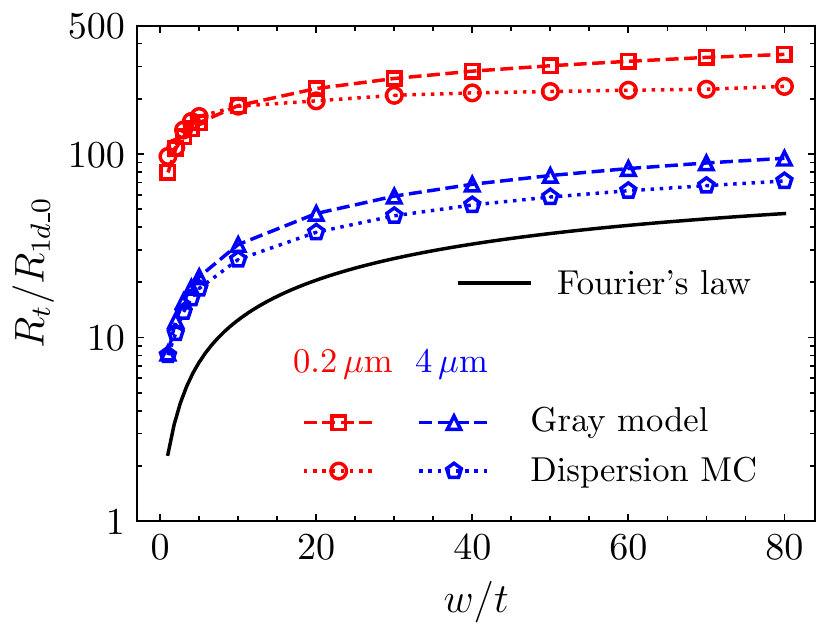}
	\caption{Dimensionless total thermal resistance of GaN as a function of $w/t$ predicted by the model and MC simulations, with different channel thicknesses, $w_g/w=0.01$.}
	\label{fig:GaN_model_0}
\end{figure}

Fig. \ref{fig:GaN_model_0} compares the values of dimensionless total thermal resistance of GaN calculated by dispersion MC and the semiempirical thermal resistance model. The predictions of Fourier's law-based model are also given for comparison. The results with different $w_g / w$ and of other materials are not exhibited since they show similar patterns. From Fig. \ref{fig:GaN_model_0}, it can be seen that both the model predictions and MC simulations are much higher than the results predicted based on Fourier's law, which indicates that the ballistic effects significantly increase the thermal resistance.

However, it can be noted that there were nonnegligible deviations between the dimensionless total thermal resistance predicted by the model and dispersion MC. For GaN in particular, the maximum relative deviation achieved 27\% with $t=\SI{4.1}{\um}$, and exceeded 33\% with $t=\SI{0.2}{\um}$. As shown in Fig. \ref{fig:GaN_model_0}, the deviations could be divided into two categories. When $w/t$ was small, the model predicted thermal resistance tended to be larger than those of dispersion MC. Moreover, with increasing $w/t$, both the dimensionless thermal resistance predicted by the model and dispersion MC increased and reached a plateau value. In the plateau region, the model predictions were lower than the MC results.

The deviations indicate that the gray model is insufficient to reflect the influences of the wide span of the phonon spectrum. As pointed out in Section \ref{subsec:model}, two kinds of ballistic effects exist in the system: the cross-plane ballistic effect and the ballistic effect with $w_g$ comparable with MFP, whose strength can be characterized by $Kn_t$ and $Kn_w$, respectively. To determine the phonon dispersion dependence of the deviations, the key issue is to separate the impacts of different ballistic effects. In the following, we first focused on the deviations related to the cross-plane ballistic part. Then, by offseting the discrepancy caused by the cross-plane ballistic effect, the $Kn_w$ dependence of the deviations from the ballistic effect with $w_g$ comparable with MFP was illustrated.

\subsection{Cross-Plane Ballistic Effect}

From equation \eqref{eq:depressed_mfp}, it can be found that in the gray model, different ballistic effects are embodied in different terms of the modified MFPs. By substituting $\left(1+A_{w}\left(w_{g} / w, w / t\right) K n_{w_{-} \omega, j}\right)$ with $r_{w}$ directly calculated from dispersion MC, as shown in equation \eqref{eq:model_1},
\begin{equation}
	l_{j, m,1} = \frac{l_{0, j}}{\left(1+\frac{2}{3} K n_{t_{-} \omega, j}\right)r_{w,\text{dispersion}}},
	\label{eq:model_1}
\end{equation}
the differences caused by the ballistic effect with $w_g$ comparable with MFP were compensated, and the remaining deviations should be attributed to the cross-plane ballistic effect. Then, the effective thermal conductivity was calculated with the newly modified MFPs $l_{j,m,1}$ using equation \eqref{eq:conductivity}, and the total thermal resistance could be evaluated.

\begin{figure}[hbtp]
	\centering
	\includegraphics[width=0.8\linewidth]{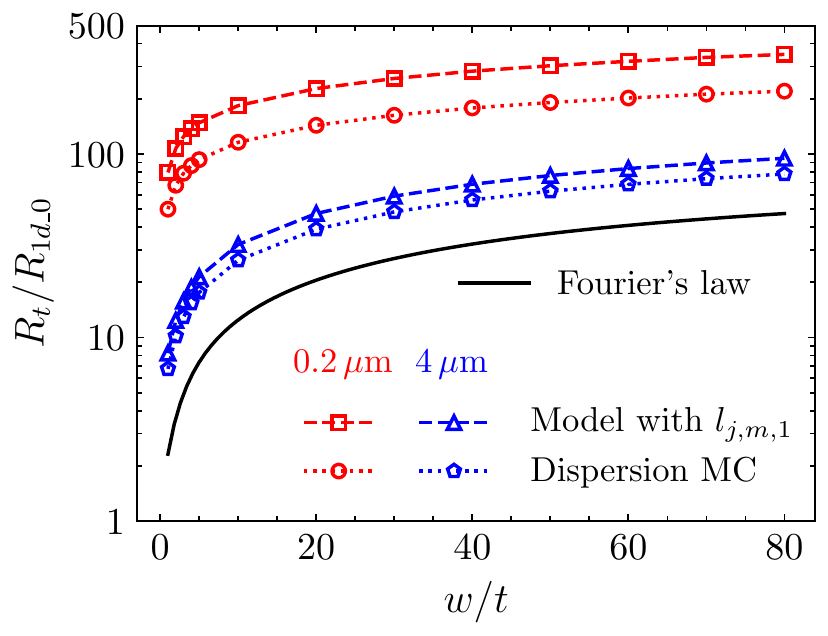}
	\caption{Dimensionless total thermal resistance of GaN as a function of $w/t$ predicted by the model based on $l_{j,m,1}$ and MC simulations, with different channel thicknesses, $w_g/w=0.01$.}
	\label{fig:GaN_model_1}
\end{figure}

Fig. \ref{fig:GaN_model_1} shows the dimensionless total thermal resistance predicted by dispersion MC and the model based on $l_{j, m, 1}$. In this case, the model's predictions were always lower than the MC results. By comparing Fig. \ref{fig:GaN_model_1} with Fig. \ref{fig:GaN_model_0}, it was found that the anomalous tendency of the model predicted thermal resistance with small $w/t$  disappeared. This indicates that the deviations of the original model in this region were mainly caused by the ballistic effect with $w_g$ comparable with MFP. In addition, as shown in Fig. \ref{fig:GaN_model_1}, the deviations increased significantly as $t$ decreased from \SI{4.1}{\mu m} to \SI{0.2}{\mu m}. This suggests that the gray model's capability to reflect the cross-plane ballistic effect was further impaired with decreasing channel thickness. To depict the channel thickness dependence of the insufficiency, an appropriate average MFP must be defined to represent the entire phonon spectrum of different materials. 

Li \etal \cite{li2020ballistic} compared different approaches to calculate the average MFP. It was found that extracting the average MFP from the fitting of the size-dependent effective thermal conductivity by a gray-medium model could well reflect the mode-dependent size effects. As shown in equation \eqref{eq:average_mfp}, the sum of the deviations with different channel thicknesses was defined as $\mathscr{L}(l_{\text{ave}})$. By employing Majumdar's model for cross-plane heat conduction \cite{majumdar1993microscale} to calculate the effective thermal conductivity and minimize $\mathscr{L}(l_{\text{ave}})$, the average MFP of different materials could be evaluated.

\begin{equation}
	\begin{aligned}
		\mathscr{L}(l_{\text{ave}}) = \sum_t & \left| \frac{1}{3} \sum_{j} \int_{0}^{\omega_{j}} \hbar \omega \frac{\partial f_{0}}{\partial T} \operatorname{DOS}_{j}(\omega) v_{g \omega j} \frac{1}{1 + \frac{4}{3}\frac{l_j}{t}} \mathrm{d} \omega \right. \\
		                                     & \left. - \frac{1}{1 + \frac{4}{3} \frac{l_{\text{ave}}}{t}} \right|^2                                                                                                                                           \\
		\label{eq:average_mfp}
	\end{aligned}
\end{equation}

The extracted average MFPs of different materials are listd in Table \ref{tab:average_mfp}. With the average MFP, the average Knudsen numbers $Kn_t$ and $Kn_w$ could then be defined as equation \eqref{eq:knudsen_t} and \eqref{eq:knudsen_w}. To clarify the $Kn_t$ dependence of the deviations between the predictions of the model and dispersion MC, a thermal resistance ratio $r_{t}=R_{\text{MC}} / R_{l\_j,m,1}$ was introduced, where $R_{\text{MC}}$ and $R_{l\_j,m,1}$ were the total thermal resistance predicted by dispersion MC and the model based on $l_{j,m,1}$, respectively. The thermal spreading effect and the ballistic effect with $w_g$ comparable with MFP were cancaled in this ratio, and thus, the ratio should reflect the cross-plane ballistic effect on thermal resistance. 

\begin{table}[htbp]
	\centering
	\caption{The average phonon MFPs of different semiconductor materials.}
	\label{tab:average_mfp}
	\scalebox{1}{
		\begin{tabular}{@{}cc@{}}
			\toprule
			Material           & Average MFP (\si{\nm}) \\ \midrule
			GaN                & 1612.3                 \\
			AlN                & 3401.4                 \\
			SiC                & 2506.9                 \\
			\ce{\beta-Ga_2O_3} & 450.7                  \\ \bottomrule
		\end{tabular}
	}
\end{table}

\begin{figure}[htbp]
	\centering
	\includegraphics[width=0.8\linewidth]{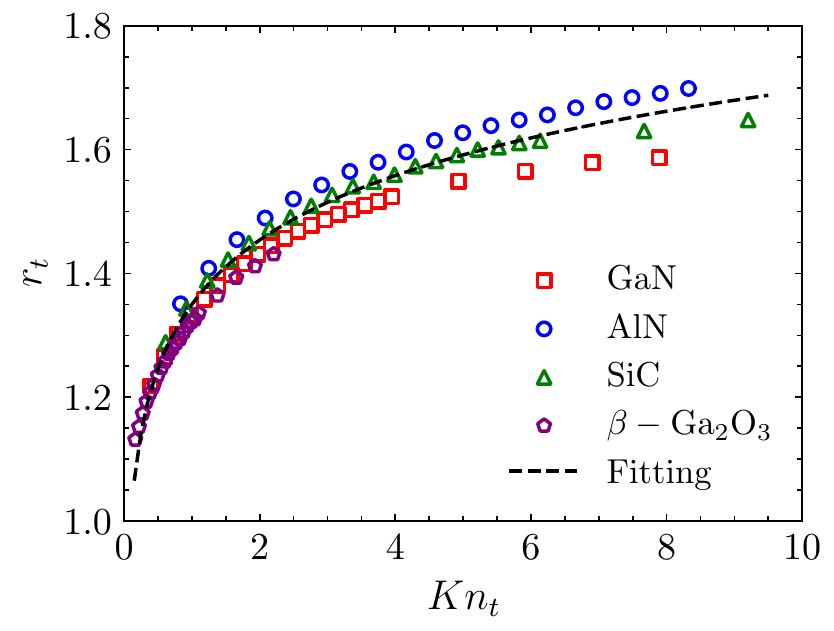}
	\caption{Thermal resistance ratio $r_{t}$ as a function of $Kn_t$ of different materials.}
	\label{fig:r_knt}
\end{figure}

Fig. \ref{fig:r_knt} shows the thermal resistance ratio varying with the average $Kn_t$ of different materials. It was found that $r_{t}$ was always larger than 1 and increased with increasing $Kn_t$. It more clearly shows that the model underestimated the cross-plane ballistic effect and that the underestimation increased with the increasing Knudsen number. The underestimation was mainly because the differential approximation was only valid for small Knudsen numbers \cite{hua2016ballistic}, and its accuracy continued to degrade with increasing $Kn_t$. As shown in \ref{fig:r_knt}, when $Kn_t$ was small, $r_{t}$ increased rapidly with $Kn_t$ since the differential approximation became inadequate for more phonons. As $Kn_t$ was large enough, the gray model was inapplicable for nearly all phonons, and the deviations gently increased with increasing $Kn_t$. From Fig. \ref{fig:r_knt}, it was found that a logarithmic linear function could well depict this $Kn_t$ dependence of $r_{t}$,
\begin{equation}
	r_t = 0.15\ln (Kn_t) + 1.35,
\end{equation}
and the fitted values coincided well with the simulated results. Specifically, as shown in Fig. \ref{fig:r_knt}, the $Kn_t$ dependences of different materials were nearly the same, which demonstrates that the way to extract the average MFP was suitable to characterize the overall size effects of the total phonon spectrum.

\subsection{Ballistic Effect with $w_g$ comparable with MFP}

By introducing $r_{t}$ to the original modified MFPs, as shown in equation \eqref{eq:model_2},
\begin{equation}
	l_{j, m, 2} = \frac{l_{0, j}}{\left(1+A_{w} K n_{w_{-} \omega, j}\right)\left(1+\frac{2}{3} K n_{t_{-} \omega, j}\right)r_t},
	\label{eq:model_2}
\end{equation}
the deviations caused by the cross-plane ballistic effect could be compensated. Thus, the remaining deviations should be attributed to the ballistic effect with $w_g$ comparable with MFP. Then, the effective thermal conductivity was be calculated with the newly modified MFPs $l_{j, m, 2}$ using equation \eqref{eq:conductivity}, and the total thermal resistance could be evaluated.

\begin{figure}[htbp]
	\centering
	\includegraphics[width=0.8\linewidth]{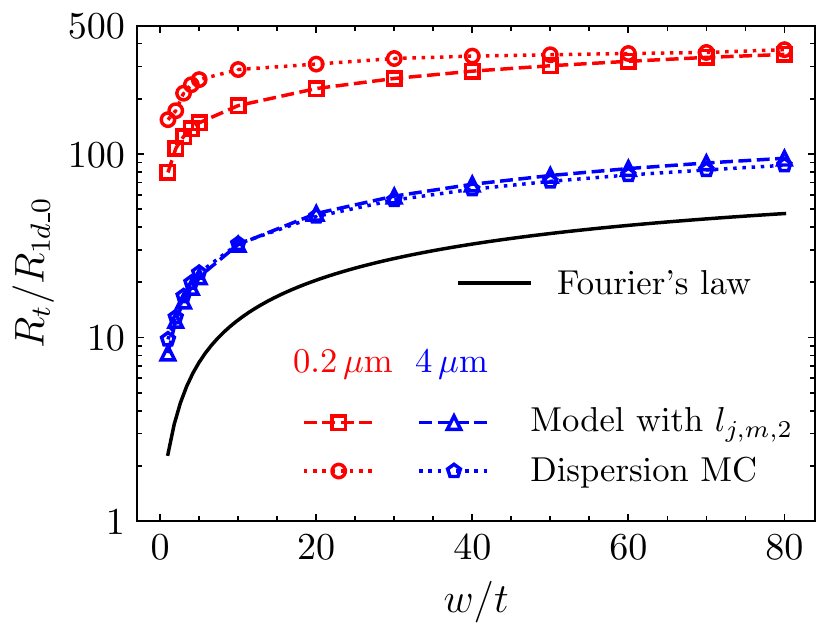}
	\caption{Dimensionless total thermal resistance of GaN as a function of $w/t$ predicted by the model based on $l_{j,m,2}$ and MC simulations, with different channel thicknesses, $w_g/w=0.01$.}
	\label{fig:fig:GaN_model_1}
\end{figure}

Fig. \ref{fig:fig:GaN_model_1} shows the dimensionless total thermal resistance of GaN predicted by dispersion MC and the model based on $l_{j, m, 2}$. It can be seen that with $t=\SI{4.1}{\um}$, the model predictions agreed well with the MC results. This means that the deviations at large channel thicknesses were mainly due to the model's underestimation of the cross-plane ballistic effect. In the case of $t=\SI{0.2}{\mu m}$, when $w/t$ is large, the model's predictions coincided well with the MC results, but the discrepancy increased significantly with decreasing $w/t$.

To clarify the $Kn_w$ dependence of this discrepancy, another thermal resistance ratio $r_{w}=R_{\text{MC}} / R_{l\_j,m,2}$ was introduced, where $R_{\text{MC}}$ and $R_{l\_j,m,2}$ were the total thermal resistance predicted by dispersion MC and the model based on $l_{j, m, 2}$, respectively. The thermal spreading effect and the cross-plane ballistic effect were cancelled in this ratio, and thus, the ratio could reflect the influences of the ballistic effect with $w_g$ comparable with MFP on the thermal resistance.

\begin{figure}[hbtp]
	\centering
	\includegraphics[width=0.8\linewidth]{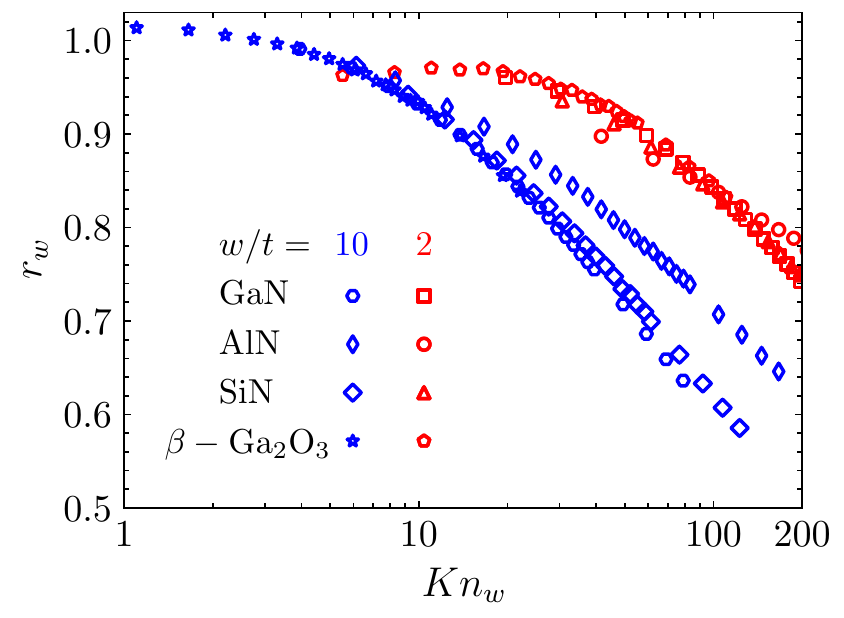}
	\caption{Thermal resistance ratio $r_{w}$ as a function of $Kn_w$ of different materials, with $w/t=2$ and $10$.}
	\label{fig:r_knw}
\end{figure}

\begin{figure}[hbtp]
	\centering
	\includegraphics[width=0.8\linewidth]{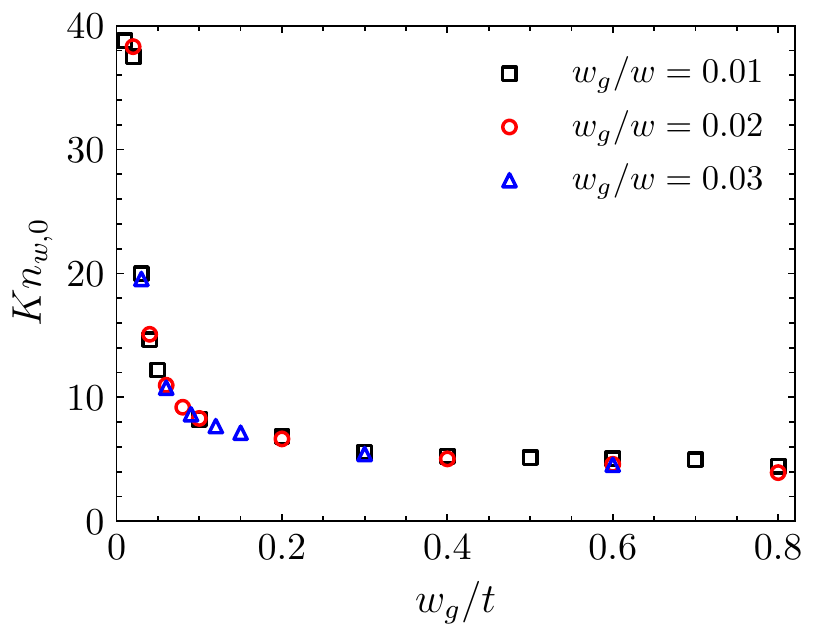}
	\caption{$Kn_{w,0}$ varying with $w_g / t$ with $w_g/w=0.01,0.02,\text{and } 0.03$.}
	\label{fig:threshold_knw}
\end{figure}

Fig. \ref{fig:r_knw} shows the thermal resistance ratio varying with the average $Kn_w$. It was found that $r_{w}$ stayed around unity with small $Kn_w$. After a threshold value $Kn_{w, 0}$, $r_{w}$ started to decrease with $Kn_w$ almost log-linearly, which was similar to the $Kn_t$ dependence of $r_{t}$. The results indicate that the fitted linear function $1 + A_wKn_w$ could well characterize the ballistic effect with $w_g$ comparable with MFP at small $Kn_w$, and its range of application was wider than the gray model for cross-plane heat conduction. The slope of this log-linear relation was nearly the same with different geometries, whereas the threshold value $Kn_{w,0}$ depended on the geometrical parameters since $A_w$ was a function of $w_g/w$ and $w/t$. To examine the influences of the geometric parameters, a log-linear function was proposed and fitted to characterize the $Kn_w$ dependence of $r_{w}$,
\begin{equation}
	r_w = -0.17 \ln (Kn_w - Kn_{w, 0}).
\end{equation}
From Fig. \ref{fig:threshold_knw}, it was found that $Kn_{w, 0}$ was only a function of $w_g/t$. As $w_g/t$ increased, $Kn_{w, 0}$ dropped rapidly and approached a plateau value. This implies that for most practical devices with large $w_g/t$, $Kn_{w, 0}$ can be approximated with a constant.

\subsection{Revised Thermal Resistance Model}

We analyzed the deviations of the total thermal resistance predicted by dispersion MC and the semiempirical thermal resistance model. Two material independent correction factors $r_{t}$ and $r_{w}$, were introduced to the model,
\begin{equation}
	l_{j, m, r} = \frac{l_{0, j}}{\left(1+A_{w} K n_{w_{-} \omega, j}\right)\left(1+\frac{2}{3} K n_{t_{-} \omega, j}\right)r_t r_w},
	\label{eq:model_3}
\end{equation}
to compensate for its insufficiency to reflect the cross-plane ballistic effect and the ballistic effect with $w_g$ comparable with MFP when considering phonon dispersion, respectively. Although this work was focused on the thermal spreading process of WBG semiconductors, our results can be extended to gray models for other systems \cite{hua2016ballistic} or to other materials.

\begin{figure}[hbtp]
	\centering
	\includegraphics[width=0.8\linewidth]{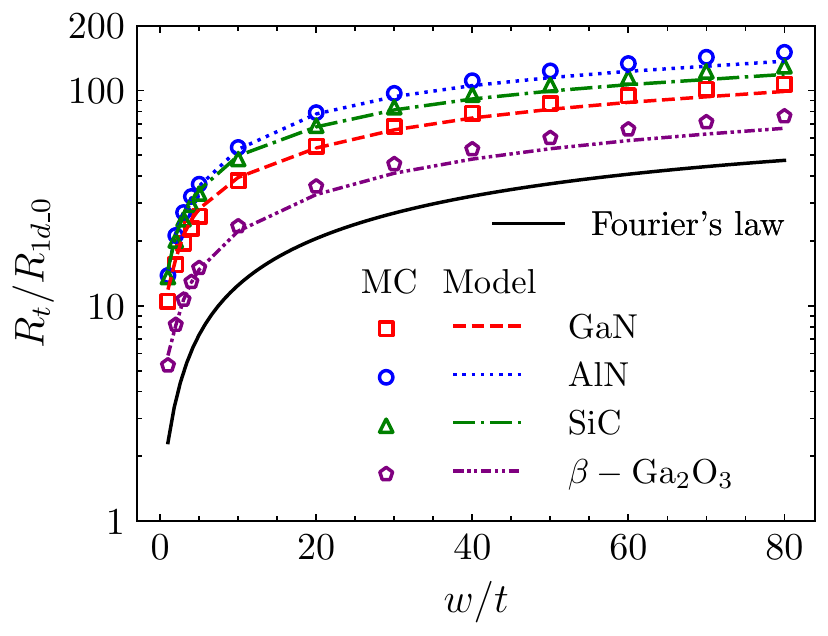}
	\caption{Dimensionless total thermal resistance of different semiconductors as a function of $w/t$ predicted by the revised model and MC simulations, with $t = \SI{2.7}{\um}$, $w_g/w=0.01$.}
	\label{fig:total_semiconductors}
\end{figure}

Fig \ref{fig:total_semiconductors} shows the dimensionless total thermal resistance predicted by dispersion MC and the revised model of different materials with $t=\SI{2.7}{\um}$. The results of Fourier's law-based model were also given for comparison. As shown in Fig. \ref{fig:total_semiconductors}, Fourier's law significantly underestimated the thermal resistance, which underscored the importance of the ballistic effects. Additionally, the separation of the curves of different materials suggests that different effective thermal conductivities should be used to reflect the influences of phonon dispersion of different materials. The predictions of the revised model coincided well with the MC simulations.

\begin{figure}[htbp]
	\captionsetup[subfloat]{farskip=0pt}
	\centering
	\subfloat[]{\includegraphics[width=0.65\linewidth]{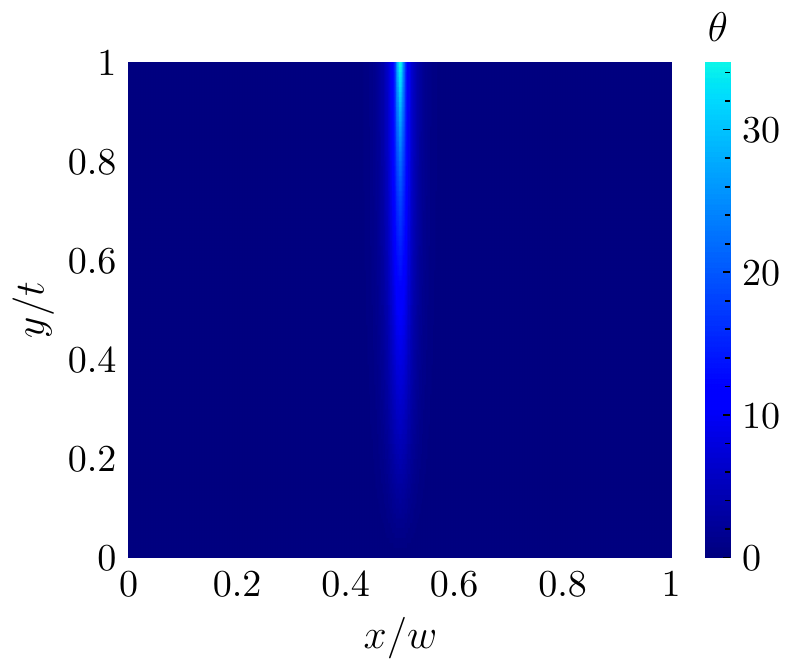}}\\
	\subfloat[]{\includegraphics[width=0.65\linewidth]{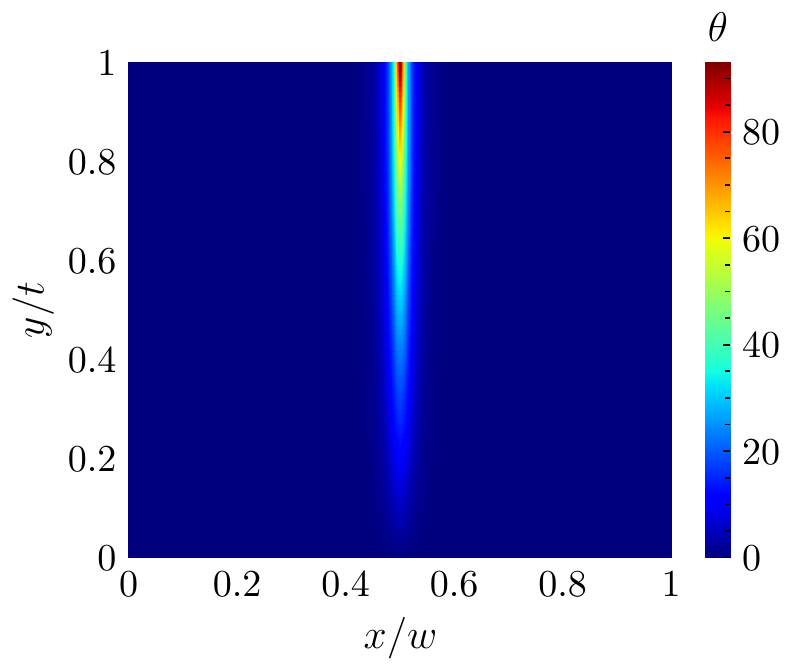}}\\
	\subfloat[]{\includegraphics[width=0.65\linewidth]{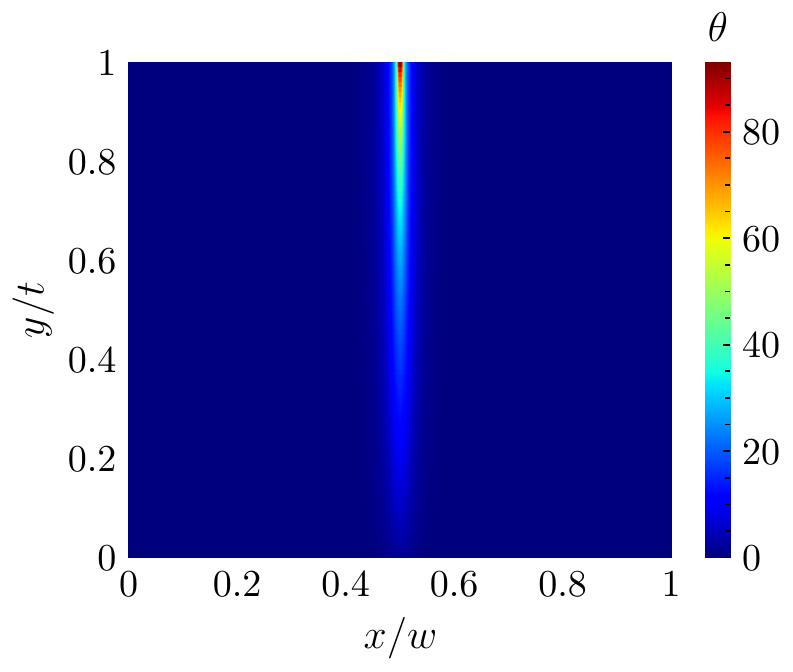}}\\
	\caption{Dimensionless temperature distributions with $w/t=40$ and $w_g/w=0.01$ of GaN, $t = \SI{2}{\um}$, predicted by (a) FEM with $k_{\text{bulk}}$, (b) FEM with $k_{\text{eff}}$, and (c) dispersion MC.}
	\label{fig:temp_field}
\end{figure}

\begin{figure}[htbp]
	\centering
	\includegraphics[width=0.8\linewidth]{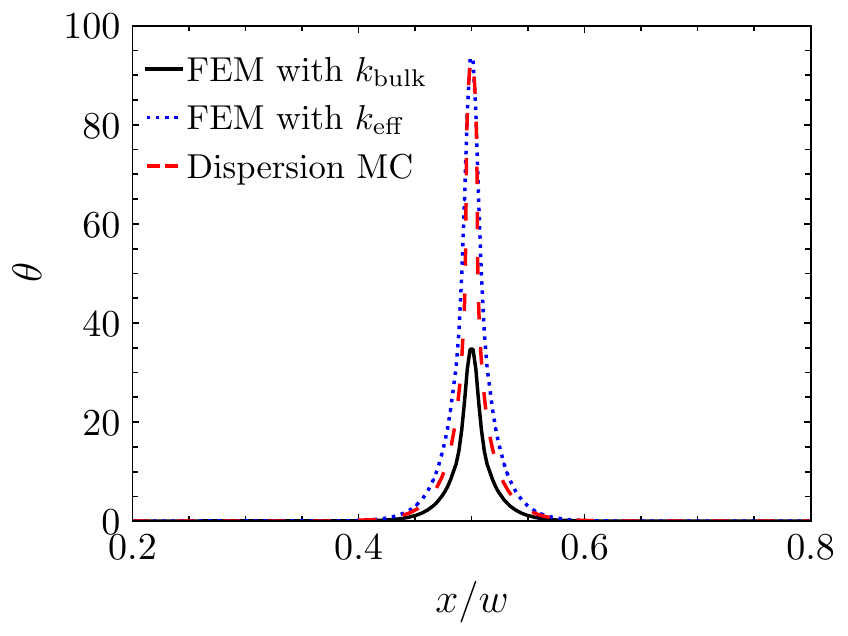}
	\caption{Dimensionless temperature distributions along the heat source plane in the system shown in Fig.\ref{fig:temp_field}.}
	\label{fig:surface_temp}
\end{figure}

Fig. \ref{fig:temp_field} shows the dimensionless temperature distributions predicted by the FEM with the bulk thermal conductivity and the effective thermal conductivity, respectively. The MC simulation result was also shown. The dimensionless temperature, $\theta$, was defined as $\theta = \Delta T / (QR_{1d\_0})$. As shown in Fig. \ref{fig:temp_field}(a), the predictions of FEM with $k_{\text{bulk}}$ significantly underestimated the peak value of dimensionless temperatures. FEM with $k_{\text{eff}}$ could give nearly the same juncture temperature as MC simulations, despite the differences in the configuration of the dimensionless temperature profile due to the intrinsic limits of Fourier's law-based methods.

Fig. \ref{fig:surface_temp} further compares the dimensionless temperature distributions along the heat source plane. It more clearly shows that the use of bulk thermal conductivities significantly underestimated the junction temperature rise. The peak dimensionless temperature rise predicted by dispersion MC was nearly 2.7 times that from FEM with bulk thermal conductivity. Well-agreed dimensionless temperature distributions predicted by the FEM with the effective thermal conductivity and MC simulations further demonstrate the effectiveness of the model.

\section{Conclusion}

We inspected the semiempirical thermal resistance model in the ballistic-diffusive regime using the phonon MC method with phonon dispersion of several typical WBG semiconductor materials. It was found that the gray model was insufficient to reflect the strong ballistic effects caused by the wide phonon MFP distributions. Two material-independent correction factors $r_{t}$ and $r_{w}$ were introduced to the model to compensate for the discrepancies caused by the cross-plane ballistic effect and the ballistic effect with $w_g$ comparable with MFP, respectively. The revised model can well address the thermal spreading process, ballistic effects, and the influences of phonon dispersion of different materials. Additionally, the model is easily coupled with widely used FEM-based thermal analysis. It was found that the FEM with the effective thermal conductivity predicted by the model can give nearly the same juncture temperature as the MC simulations, which was much higher than the results predicted with bulk thermal conductivities. This paper can provide a clearer understanding of the influences of phonon dispersion on the thermal transport process, and the model can be useful for the thermal management of WBG or UWBG semiconductor devices. Additionally, the same analysis process and model construction methodology are expected to be applied in more complex and practical device situations, such as uneven internal heat source distributions \cite{hua2016effective}, and coupled with electro-thermal device simulations\cite{choi2021perspective}. 

\bibliographystyle{IEEEtran}
\bibliography{manuscript.bib}

\begin{thebibliography}{10}
\providecommand{\url}[1]{#1}
\csname url@samestyle\endcsname
\providecommand{\newblock}{\relax}
\providecommand{\bibinfo}[2]{#2}
\providecommand{\BIBentrySTDinterwordspacing}{\spaceskip=0pt\relax}
\providecommand{\BIBentryALTinterwordstretchfactor}{4}
\providecommand{\BIBentryALTinterwordspacing}{\spaceskip=\fontdimen2\font plus
\BIBentryALTinterwordstretchfactor\fontdimen3\font minus
  \fontdimen4\font\relax}
\providecommand{\BIBforeignlanguage}[2]{{%
\expandafter\ifx\csname l@#1\endcsname\relax
\typeout{** WARNING: IEEEtran.bst: No hyphenation pattern has been}%
\typeout{** loaded for the language `#1'. Using the pattern for}%
\typeout{** the default language instead.}%
\else
\language=\csname l@#1\endcsname
\fi
#2}}
\providecommand{\BIBdecl}{\relax}
\BIBdecl

\bibitem{mishra2002algan}
U.~K. Mishra, P.~Parikh, and Y.-F. Wu, ``{{\ce{AlGaN/GaN}}} {HEMTs}-an overview
  of device operation and applications,'' \emph{Proceedings of the IEEE},
  vol.~90, no.~6, pp. 1022--1031, 2002.

\bibitem{zeng2018comprehensive}
F.~Zeng, J.~X. An, G.~Zhou, W.~Li, H.~Wang, T.~Duan, L.~Jiang, and H.~Yu, ``A
  comprehensive review of recent progress on {{\ce{GaN}}} high electron
  mobility transistors: Devices, fabrication and reliability,''
  \emph{Electronics}, vol.~7, no.~12, p. 377, 2018.

\bibitem{sriram2009high}
S.~Sriram, H.~Hagleitner, D.~Namishia, T.~Alcorn, T.~Smith, and B.~Pulz,
  ``High-gain {{\ce{SiC}}} {MESFETs} using source-connected field plates,''
  \emph{IEEE Electron Device Letters}, vol.~30, no.~9, pp. 952--953, 2009.

\bibitem{zhu2019improved}
S.~Zhu, H.~Jia, X.~Wang, Y.~Liang, Y.~Tong, T.~Li, and Y.~Yang, ``Improved mrd
  {{\ce{4H-SiC}}} {MESFET} with high power added efficiency,''
  \emph{Micromachines}, vol.~10, no.~7, p. 479, 2019.

\bibitem{higashiwaki2012gallium}
M.~Higashiwaki, K.~Sasaki, A.~Kuramata, T.~Masui, and S.~Yamakoshi, ``Gallium
  oxide ({{\ce{Ga_2O_3}}}) metal-semiconductor field-effect transistors on
  single-crystal {{\ce{\beta-Ga_2O_3}}} (010) substrates,'' \emph{Applied
  Physics Letters}, vol. 100, no.~1, p. 013504, 2012.

\bibitem{wong2015field}
M.~H. Wong, K.~Sasaki, A.~Kuramata, S.~Yamakoshi, and M.~Higashiwaki,
  ``Field-plated {{\ce{Ga_2O_3}}} {MOSFETs} with a breakdown voltage of over
  750 v,'' \emph{IEEE Electron Device Letters}, vol.~37, no.~2, pp. 212--215,
  2015.

\bibitem{wang2021progresses}
C.~Wang, J.~Zhang, S.~Xu, C.~Zhang, Q.~Feng, Y.~Zhang, J.~Ning, S.~Zhao,
  H.~Zhou, and Y.~Hao, ``Progresses in state-of-the-art technologies of
  {{\ce{\beta-Ga_2O_3}}} devices,'' \emph{Journal of Physics D: Applied
  Physics}, 2021.

\bibitem{wu2007transient}
Y.-R. Wu and J.~Singh, ``Transient study of self-heating effects in
  {{\ce{AlGaN/GaN}}} {HFETs}: Consequence of carrier velocities, temperature,
  and device performance,'' \emph{Journal of Applied Physics}, vol. 101,
  no.~11, p. 113712, 2007.

\bibitem{cheng2019significantly}
Z.~Cheng, N.~Tanen, C.~Chang, J.~Shi, J.~McCandless, D.~Muller, D.~Jena, H.~G.
  Xing, and S.~Graham, ``Significantly reduced thermal conductivity in
  {{\ce{\beta-(Al_{0.1}Ga_{0.9})_2O_3 / Ga_2O_3} }}superlattices,''
  \emph{Applied Physics Letters}, vol. 115, no.~9, p. 092105, 2019.

\bibitem{liu2020machine}
Y.-B. Liu, J.-Y. Yang, G.-M. Xin, L.-H. Liu, G.~Cs{\'a}nyi, and B.-Y. Cao,
  ``Machine learning interatomic potential developed for molecular simulations
  on thermal properties of {{\ce{\beta-Ga_2O_3}}},'' \emph{The Journal of
  Chemical Physics}, vol. 153, no.~14, p. 144501, 2020.

\bibitem{waltereit2012influence}
P.~Waltereit, W.~Bronner, M.~Musser, F.~v. Raay, M.~Dammann, M.~C{\"a}sar,
  S.~M{\"u}ller, L.~Kirste, K.~K{\"o}hler, R.~Quay \emph{et~al.}, ``Influence
  of {{\ce{AlGaN}}} barrier thickness on electrical and device properties in
  {{\ce{Al_{0.14}Ga_{0.86}N/GaN}}} high electron mobility transistor
  structures,'' \emph{Journal of Applied Physics}, vol. 112, no.~5, p. 053718,
  2012.

\bibitem{bagnall2013device}
K.~R. Bagnall, ``Device-level thermal analysis of gan-based electronics,''
  Ph.D. dissertation, Massachusetts Institute of Technology, 2013.

\bibitem{tang2020thermal}
D.-S. Tang, G.-Z. Qin, M.~Hu, and B.-Y. Cao, ``Thermal transport properties of
  gan with biaxial strain and electron-phonon coupling,'' \emph{Journal of
  Applied Physics}, vol. 127, no.~3, p. 035102, 2020.

\bibitem{tang2021phonon}
D.-S. Tang and B.-Y. Cao, ``Phonon thermal transport properties of gan with
  symmetry-breaking and lattice deformation induced by the electric field,''
  \emph{International Journal of Heat and Mass Transfer}, vol. 179, p. 121659,
  2021.

\bibitem{chatterjee2019device}
B.~Chatterjee, K.~Zeng, C.~D. Nordquist, U.~Singisetti, and S.~Choi,
  ``Device-level thermal management of gallium oxide field-effect
  transistors,'' \emph{IEEE Transactions on Components, Packaging and
  Manufacturing Technology}, vol.~9, no.~12, pp. 2352--2365, 2019.

\bibitem{chatterjee2020nanoscale}
B.~Chatterjee, C.~Dundar, T.~E. Beechem, E.~Heller, D.~Kendig, H.~Kim,
  N.~Donmezer, and S.~Choi, ``Nanoscale electro-thermal interactions in
  {{\ce{AlGaN/GaN}}} high electron mobility transistors,'' \emph{Journal of
  Applied Physics}, vol. 127, no.~4, p. 044502, 2020.

\bibitem{sarua2007thermal}
A.~Sarua, H.~Ji, K.~Hilton, D.~Wallis, M.~J. Uren, T.~Martin, and M.~Kuball,
  ``Thermal boundary resistance between {{\ce{GaN}}} and substrate in
  {{\ce{AlGaN/GaN}}} electronic devices,'' \emph{IEEE Transactions on electron
  devices}, vol.~54, no.~12, pp. 3152--3158, 2007.

\bibitem{razavi2016review}
M.~Razavi, Y.~Muzychka, and S.~Kocabiyik, ``Review of advances in thermal
  spreading resistance problems,'' \emph{Journal of Thermophysics and Heat
  Transfer}, vol.~30, no.~4, pp. 863--879, 2016.

\bibitem{muzychka2013thermal}
Y.~S. Muzychka, K.~R. Bagnall, and E.~N. Wang, ``Thermal spreading resistance
  and heat source temperature in compound orthotropic systems with interfacial
  resistance,'' \emph{IEEE Transactions on Components, Packaging and
  Manufacturing Technology}, vol.~3, no.~11, pp. 1826--1841, 2013.

\bibitem{garven2009simulation}
M.~Garven and J.~P. Calame, ``Simulation and optimization of gate temperatures
  in {GaN-on-SiC} monolithic microwave integrated circuits,'' \emph{IEEE
  Transactions on Components and Packaging Technologies}, vol.~32, no.~1, pp.
  63--72, 2009.

\bibitem{orouji2011novel}
A.~A. Orouji and A.~Aminbeidokhti, ``A novel double-recessed {4H-SiC} mesfet
  with partly undoped space region,'' \emph{Superlattices and Microstructures},
  vol.~50, no.~6, pp. 680--690, 2011.

\bibitem{chen2021non}
G.~Chen, ``Non-fourier phonon heat conduction at the microscale and
  nanoscale,'' \emph{Nature Reviews Physics}, pp. 1--15, 2021.

\bibitem{hua2016ballistic}
Y.-C. Hua and B.-Y. Cao, ``Ballistic-diffusive heat conduction in
  multiply-constrained nanostructures,'' \emph{International Journal of Thermal
  Sciences}, vol. 101, pp. 126--132, 2016.

\bibitem{hua2019thermal}
Y.-C. Hua, H.-L. Li, and B.-Y. Cao, ``Thermal spreading resistance in
  ballistic-diffusive regime for {GaN HEMTs},'' \emph{IEEE Transactions on
  Electron Devices}, vol.~66, no.~8, pp. 3296--3301, 2019.

\bibitem{bao2018review}
H.~Bao, J.~Chen, X.~Gu, and B.~Cao, ``A review of simulation methods in
  micro/nanoscale heat conduction,'' \emph{ES Energy \& Environment}, vol.~1,
  no.~34, pp. 16--55, 2018.

\bibitem{sadi2010monte}
T.~Sadi and R.~W. Kelsall, ``Monte carlo study of the electrothermal phenomenon
  in silicon-on-insulator and silicon-germanium-on-insulator metal-oxide
  field-effect transistors,'' \emph{Journal of Applied Physics}, vol. 107,
  no.~6, p. 064506, 2010.

\bibitem{hao2017hybrid}
Q.~Hao, H.~Zhao, and Y.~Xiao, ``A hybrid simulation technique for
  electrothermal studies of two-dimensional {GaN-on-SiC} high electron mobility
  transistors,'' \emph{Journal of Applied Physics}, vol. 121, no.~20, p.
  204501, 2017.

\bibitem{hao2018hybrid}
Q.~Hao, H.~Zhao, Y.~Xiao, Q.~Wang, and X.~Wang, ``Hybrid electrothermal
  simulation of a 3-d fin-shaped field-effect transistor based on {GaN}
  nanowires,'' \emph{IEEE Transactions on Electron Devices}, vol.~65, no.~3,
  pp. 921--927, 2018.

\bibitem{li2020ballistic}
H.-L. Li, J.~Shiomi, and B.-Y. Cao, ``Ballistic-diffusive heat conduction in
  thin films by phonon monte carlo method: Gray medium approximation versus
  phonon dispersion,'' \emph{Journal of Heat Transfer}, vol. 142, no.~11, p.
  112502, 2020.

\bibitem{wang2013simulation}
A.~Wang, M.~Tadjer, and F.~Calle, ``Simulation of thermal management in
  {AlGaN/GaN HEMTs} with integrated diamond heat spreaders,''
  \emph{Semiconductor science and technology}, vol.~28, no.~5, p. 055010, 2013.

\bibitem{zhang2019enhancement}
H.~Zhang, Z.~Guo, and Y.~Lu, ``Enhancement of hot spot cooling by capped
  diamond layer deposition for multifinger {AlGaN/GaN HEMTs},'' \emph{IEEE
  Transactions on Electron Devices}, vol.~67, no.~1, pp. 47--52, 2019.

\bibitem{yuan2020modeling}
C.~Yuan, Y.~Zhang, R.~Montgomery, S.~Kim, J.~Shi, A.~Mauze, T.~Itoh, J.~S.
  Speck, and S.~Graham, ``Modeling and analysis for thermal management in
  gallium oxide field-effect transistors,'' \emph{Journal of Applied Physics},
  vol. 127, no.~15, p. 154502, 2020.

\bibitem{bagnall2014analytical}
K.~R. Bagnall, Y.~S. Muzychka, and E.~N. Wang, ``Analytical solution for
  temperature rise in complex multilayer structures with discrete heat
  sources,'' \emph{IEEE Transactions on Components, Packaging and Manufacturing
  Technology}, vol.~4, no.~5, pp. 817--830, 2014.

\bibitem{freedman2013universal}
J.~P. Freedman, J.~H. Leach, E.~A. Preble, Z.~Sitar, R.~F. Davis, and J.~A.
  Malen, ``Universal phonon mean free path spectra in crystalline
  semiconductors at high temperature,'' \emph{Scientific reports}, vol.~3,
  no.~1, pp. 1--6, 2013.

\bibitem{muzychka2003thermal}
Y.~Muzychka, J.~Culham, and M.~Yovanovich, ``Thermal spreading resistance of
  eccentric heat sources on rectangular flux channels,'' \emph{J. Electron.
  Packag.}, vol. 125, no.~2, pp. 178--185, 2003.

\bibitem{hua2017slip}
Y.-C. Hua and B.-Y. Cao, ``Slip boundary conditions in ballistic--diffusive
  heat transport in nanostructures,'' \emph{Nanoscale and Microscale
  Thermophysical Engineering}, vol.~21, no.~3, pp. 159--176, 2017.

\bibitem{tang2016phonon}
D.-S. Tang, Y.-C. Hua, B.-D. Nie, and B.-Y. Cao, ``Phonon wave propagation in
  ballistic-diffusive regime,'' \emph{Journal of Applied Physics}, vol. 119,
  no.~12, p. 124301, 2016.

\bibitem{hua2014phonon}
Y.-C. Hua and B.-Y. Cao, ``Phonon ballistic-diffusive heat conduction in
  silicon nanofilms by monte carlo simulations,'' \emph{International Journal
  of Heat and Mass Transfer}, vol.~78, pp. 755--759, 2014.

\bibitem{peraud2011efficient}
J.-P.~M. P{\'e}raud and N.~G. Hadjiconstantinou, ``Efficient simulation of
  multidimensional phonon transport using energy-based variance-reduced monte
  carlo formulations,'' \emph{Physical Review B}, vol.~84, no.~20, p. 205331,
  2011.

\bibitem{chung2004role}
J.~Chung, A.~McGaughey, and M.~Kaviany, ``Role of phonon dispersion in lattice
  thermal conductivity modeling,'' \emph{J. Heat Transfer}, vol. 126, no.~3,
  pp. 376--380, 2004.

\bibitem{chen2005nanoscale}
G.~Chen, \emph{Nanoscale energy transport and conversion: a parallel treatment
  of electrons, molecules, phonons, and photons}.\hskip 1em plus 0.5em minus
  0.4em\relax Oxford university press, 2005.

\bibitem{guo2015anisotropic}
Z.~Guo, A.~Verma, X.~Wu, F.~Sun, A.~Hickman, T.~Masui, A.~Kuramata,
  M.~Higashiwaki, D.~Jena, and T.~Luo, ``Anisotropic thermal conductivity in
  single crystal $\beta$-gallium oxide,'' \emph{Applied Physics Letters}, vol.
  106, no.~11, p. 111909, 2015.

\bibitem{yan2018phonon}
Z.~Yan and S.~Kumar, ``Phonon mode contributions to thermal conductivity of
  pristine and defective {{\ce{\beta-Ga_2O_3}}},'' \emph{Physical Chemistry
  Chemical Physics}, vol.~20, no.~46, pp. 29\,236--29\,242, 2018.

\bibitem{jiang2018three}
P.~Jiang, X.~Qian, X.~Li, and R.~Yang, ``Three-dimensional anisotropic thermal
  conductivity tensor of single crystalline $\beta$-ga2o3,'' \emph{Applied
  Physics Letters}, vol. 113, no.~23, p. 232105, 2018.

\bibitem{danilchenko2007upper}
B.~Danilchenko, I.~Obukhov, T.~Paszkiewicz, S.~Wolski, and A.~Je{\.z}owski,
  ``On the upper limit of thermal conductivity gan crystals,'' \emph{Solid
  state communications}, vol. 144, no. 3-4, pp. 114--117, 2007.

\bibitem{slack1987intrinsic}
G.~A. Slack, R.~A. Tanzilli, R.~Pohl, and J.~Vandersande, ``The intrinsic
  thermal conductivity of aln,'' \emph{Journal of Physics and Chemistry of
  Solids}, vol.~48, no.~7, pp. 641--647, 1987.

\bibitem{zheng2019thermal}
Q.~Zheng, C.~Li, A.~Rai, J.~H. Leach, D.~A. Broido, and D.~G. Cahill, ``Thermal
  conductivity of gan, gan 71, and sic from 150 k to 850 k,'' \emph{Physical
  Review Materials}, vol.~3, no.~1, p. 014601, 2019.

\bibitem{majumdar1993microscale}
A.~Majumdar, ``{Microscale Heat Conduction in Dielectric Thin Films},''
  \emph{Journal of Heat Transfer}, vol. 115, no.~1, pp. 7--16, 02 1993.

\bibitem{hua2016effective}
Y.-C. Hua and B.-Y. Cao, ``The effective thermal conductivity of
  ballistic--diffusive heat conduction in nanostructures with internal heat
  source,'' \emph{International Journal of Heat and Mass Transfer}, vol.~92,
  pp. 995--1003, 2016.

\bibitem{choi2021perspective}
S.~Choi, S.~Graham, S.~Chowdhury, E.~R. Heller, M.~J. Tadjer, G.~Moreno, and
  S.~Narumanchi, ``A perspective on the electro-thermal co-design of ultra-wide
  bandgap lateral devices,'' \emph{Applied Physics Letters}, vol. 119, no.~17,
  p. 170501, 2021.

\end{thebibliography}

\end{document}